\documentclass{PoS}

\newcommand{\dm}[1]{\Delta m^2_{#1}}
\newcommand{\dcp}{\delta_{CP}}
\newcommand{\pmue}{P_{\mu e}}
\newcommand{\essnusb}{ESS$\nu$SB}

\title{Synergies and complementarities between proposed future neutrino projects}

\ShortTitle{Synergies between future neutrino projects}

\author{\speaker{Sushant K. Raut}\\
        Center for Theoretical Physics of the Universe, Institute for Basic Science (IBS), Daejeon, 34051, Korea\\
        E-mail: \email{sushant@ibs.re.kr}}

\abstract{Measuring the unknown neutrino oscillation parameters is 
one of the main aims in neutrino physics today. 
The measurement of these parameters is 
severely affected by the presence of degeneracies in the parameter
space. Various neutrino oscillation projects have been proposed to 
measure them. In this overview talk, we will discuss some of the 
proposed neutrino experiments and the synergies between them that 
can enhance their future physics reach. }

\FullConference{The 19th International Workshop on Neutrinos from Accelerators-NUFACT2017\\
		25-30 September, 2017\\
		Uppsala University, Uppsala, Sweden}

\begin{document}

\section{General framework}

The discovery of neutrino oscillations and hence of non-zero neutrino 
masses was the first sign of a deviation from the canonical Standard Model. 
Over the last two decades, neutrino oscillation physics has increasingly 
been seen as a window to new physics, in particular on models that could 
explain neutrino masses and mixing. While the mechanism of neutrino 
oscillations does not depend directly on these models, the values of the 
oscillation parameters can be useful discriminators between competing 
models. 

In the standard three-flavour framework, neutrino oscillation 
probabilities depend on three mixing angles -- $\theta_{12}$, $\theta_{13}$ 
and $\theta_{23}$, two mass-squared differences between the neutrino mass 
eigenstates -- $\dm{21}$ and $\dm{31}$, and the CP-violating phase $\dcp$. 
Solar and reactor neutrino experiments have already measured $\theta_{12}$, 
$\theta_{13}$ and $\dm{21}$ to very good precision. According to recent 
global fits~\cite{Capozzi:2017ipn}, the magnitude of 
$\dm{31}$ is known to a few percent precision but its sign (positive, called 
normal hierarchy (NH) or negative, called inverted hierarchy (IH)) is not 
known. The octant of the close-to-maximal mixing angle $\theta_{23}$ 
($<45^\circ$, lower octant (LO) or $>45^\circ$, higher octant (HO)) is 
unknown. The value of the phase $\dcp$ is also largely unknown at present. 
In addition, if extra sterile neutrino states or 
non-standard interactions (NSIs) exist in nature, they may introduce 
more parameters that are yet to be measured. 

In this overview talk, we will discuss the parameter degeneracies that 
affect the measurement of the above parameters (primarily within the 
standard oscillation framework). We will see how 
various proposed experiments can achieve these measurements, singly and 
through a synergistic combination with other facilities. (Disclaimer: The 
list of references cited in this proceedings article are chosen to highlight 
some physics points and must not be considered exhaustive.)

\section{Parameter degeneracies}

The $\nu_\mu \to \nu_e$ oscillation channel is sensitive to 
all three unknown parameters and hence the probability $\pmue$ is 
useful for understanding the reach of various beam-based and atmospheric 
neutrino experiments. These experiments measure event rates which 
depend on $\pmue$ and hence the unknown parameters. It is possible 
for multiple combinations of these unknown parameters to give the 
same oscillation probability. This is the problem of parameter 
degeneracy~\cite{Barger:2001yr}. 

For instance, the enhancement in $\pmue$ due to matter effects in case 
of NH can be offset if $\dcp$ is close to $90^\circ$. Similarly, the 
decrease in $\pmue$ due to matter effects in case 
of IH can be offset if $\dcp$ is close to $-90^\circ$. Therefore 
the combination of NH and $\dcp$ in the upper half-plane (UHP) is 
degenerate with the combination of IH and $\dcp$ in the lower 
half-plane (LHP)~\cite{Prakash:2012az,Agarwalla:2012bv}. This degeneracy is the same for antineutrinos.
For the octant, it is found that the hierarchy-octant combination of 
NH and LO is degenerate with IH and HO in case of neutrinos, while 
NH and HO is degenerate with IH and LO for antineutrinos~\cite{Agarwalla:2013ju}. All 
these features can be explained using the approximate analytic 
expressions for $\pmue$~\cite{Akhmedov:2004ny}.

\begin{figure*}
\begin{center}
%\vspace{-0.1cm}
\includegraphics[trim=0 50 0 80,clip,width=0.85\textwidth]{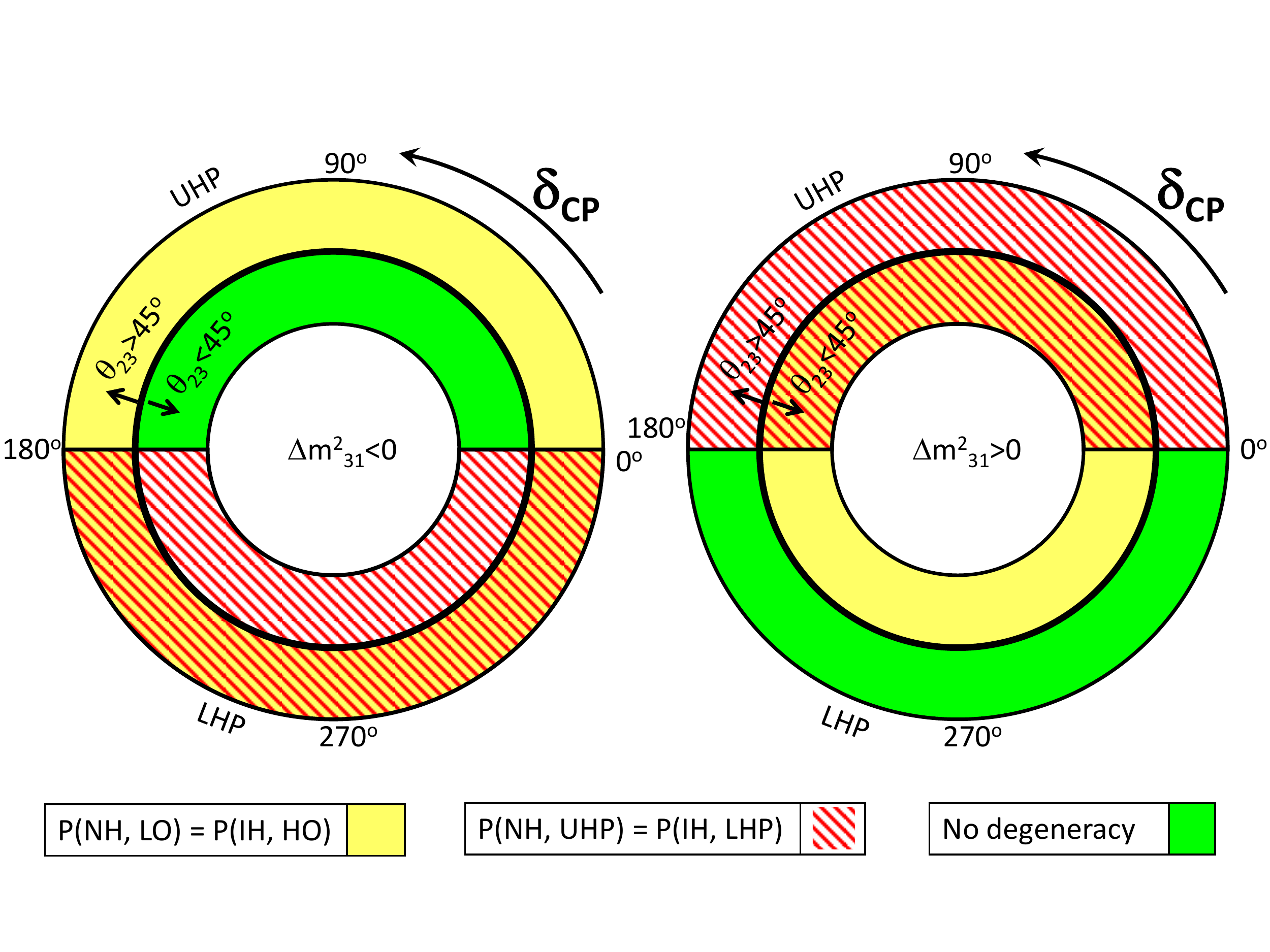}
\caption{Schematic representation of parameter degeneracies in 
three-flavour neutrino oscillations. See the text for a detailed description.}
\label{fig:nudegen}
\end{center}
\end{figure*}

\begin{figure*}
%\vspace{-5cm}
\begin{center}
\includegraphics[trim=0 50 0 80,clip,width=0.85\textwidth]{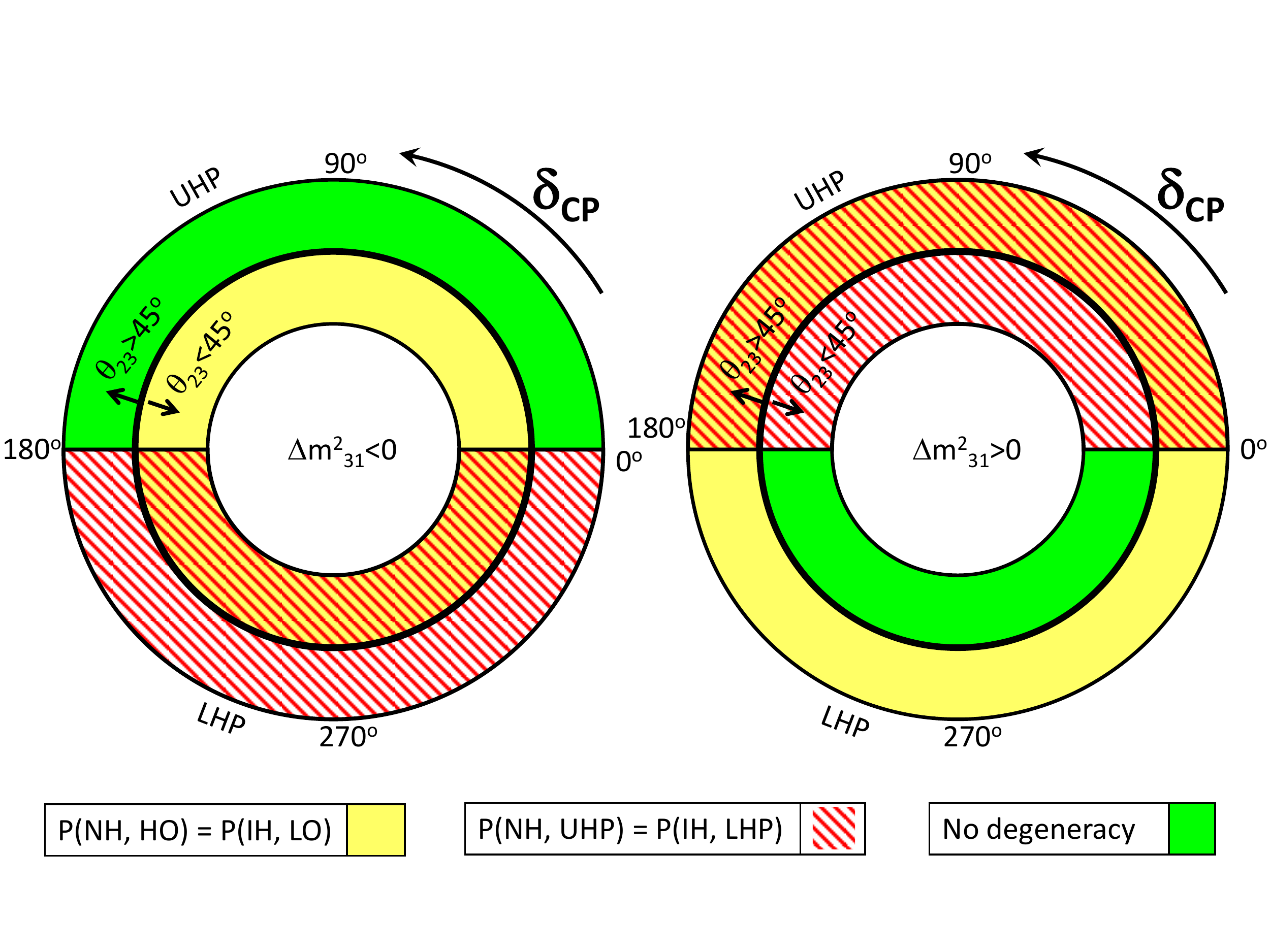}
\caption{Same as Fig.~1, but for antineutrinos.}
\label{fig:anudegen}
\end{center}
\end{figure*}

Figures~\ref{fig:nudegen} and ~\ref{fig:anudegen} represent these 
degeneracies schematically for neutrinos and antineutrinos, respectively. 
In each figure, the left(right) part represents the parameter space for 
IH(NH). Within each of these parts, the inner(outer) ring corresponds 
to the choice of octant LO(HO). In each ring, the angular direction 
gives the value of $\dcp$. The $\dcp$ parameter space is divided into 
two half-planes LHP and UHP. Thus the complete parameter space in each 
figure is divided into eight parts, corresponding to the two choices of 
hierarchy, octant and half-plane of $\dcp$. 

%-------------------------------------------------------------------------
%--------------------------------------------------------------------------
%--------------------------------------------------------------------------

Next, we shade the parts of this parameter space that are degenerate with 
each other, as indicated in the legend. The octant degeneracy is coloured 
yellow while the hierarchy-$\dcp$ degeneracy is shaded with red stripes. 
The parts of the diagram that are left unshaded are marked in green and 
are free of parameter degeneracies. If the values of parameters in 
nature lie in the green region of Fig.~\ref{fig:nudegen}, the neutrino run 
of the current generation of experiments will measure them without being 
hampered by degeneracies. If they lie in the yellow solid region, collecting 
data with antineutrinos will break the octant degeneracy since it affects 
neutrinos and antineutrinos in opposite ways (compare Figs.~\ref{fig:nudegen} and ~\ref{fig:anudegen}). 
However, if the parameters lie in the striped region, switching polarities 
does not help to lift the hierarchy degeneracy.

\section{Proposed future neutrino projects}

Some of the proposed experiments that will aim to measure the parameters 
are the long-baseline superbeam experiments DUNE~\cite{Acciarri:2015uup}, 
T2HK~\cite{t2hk}, T2HKK~\cite{Abe:2016ero} and \essnusb~\cite{ESS}, the 
medium-baseline reactor neutrino experiment JUNO~\cite{Li:2014qca}, 
the atmospheric neutrino experiments HK~\cite{hkdesignreport}, ICAL@INO~\cite{Ahmed:2015jtv} 
and Deep Core/PINGU~\cite{Aartsen:2014oha}, 
and experiments based on a muon decay source, either in-flight like 
MOMENT~\cite{moment} or at rest~\cite{Conrad:2009mh}. 

Each experiment has its characteristic baseline and energy, and hence 
has a different functional dependence on the oscillation parameters. 
Thus the degeneracies experienced by them are, in general, different. 
While all the experiments will point to the same correct set of 
parameters, they can help to exclude each others fake degenerate 
solutions. This synergy between the experiments will allow us to measure 
the oscillation parameters unambiguously, even if the individual experiments 
suffer from degeneracies~\cite{Ghosh:2013pfa}.

\section{Measurement of unknown parameters}

First, we address the issue of determining the neutrino mass 
hierarchy. As discussed before, this can be impeded by the presence of the 
hierarchy-$\dcp$ degeneracy, if the combination of parameters in nature is 
unfavourable. 
In that case, in addition to the current generation of 
experiments, we need information from an experiment 
that has (a) more matter effects (in order to break the 
degeneracy), or (b) is insensitive to $\dcp$ (so that the hierarchy can be 
measured independently), or (c) has negligible matter effects (allowing $\dcp$ to 
be measured independently, thus lifting the degeneracy). 

\noindent
(a) The degeneracy is broken with more matter effects by future long-baseline experiments such 
as DUNE (which will exclude the wrong hierarchy at at least $3\sigma$ 
even for unfavourable parameter values)~\cite{Barger:2013rha}. 
The T2HKK proposal which aims to 
locate half of the HK detector in Korea in the path of the T2HK beam 
will combine data from the shorter T2HK and longer Tokai-to-Korea baselines. 
The addition of the longer-baseline component with more matter effects will
increase the hierarchy sensitivity in the least favourable region from 
around $1\sigma$ to $6\sigma$~\cite{Abe:2016ero}. 

\noindent
(b) Experiments that are insensitive to $\dcp$ such as atmospheric neutrino 
experiments can also determine the mass hierarchy, individually or in 
combination with long-baseline experiments. This has been illustrated in, 
for example, Refs.~\cite{Ghosh:2012px} and ~\cite{Winter:2013ema} in the context of ICAL and PINGU. 
The $\dcp$-independent hierarchy sensitivity of the atmospheric neutrino 
experiment pushes the combined sensitivity of the long-baseline+atmospheric 
experiments to higher confidence levels~\cite{Barger:2013rha,Ghosh:2014rna}. 

\noindent
(c) Experiments such as the muon decay-at-rest facilities with very short baselines are capable of measuring 
$\dcp$ accurately, because there is no interference between the CP and 
matter effects. These have 
traditionally been studied in the context of CP-measurements, but can also 
improve the mass hierarchy sensitivity~\cite{Agarwalla:2017nld}. 
In spite of having negligible hierarchy sensitivity by themselves, these 
experiments help by excluding the fake degenerate CP solution of long-baseline 
experiments, demonstrating remarkable synergy between the two.

It is also worth mentioning that the JUNO experiment aims to determine the 
mass hierarchy using a completely different method, i.e. by resolving 
the difference between the solar scale 
oscillations in case of NH and IH~\cite{Li:2014qca}. The confidence level 
at which the two can be distinguished 
depends on the energy resolution of the experiment. 

Next, we discuss the measurement of $\dcp$. Once again, we can classify 
experiments into the three above categories. Long-baseline experiments 
with more matter effects can break the degeneracy between the 
unfavourable regions for NH and IH and determine the correct hierarchy, 
thus picking out the correct $\dcp$ solution. This results in an 
enhanced CP discovery potential (see for example Refs.~\cite{Barger:2013rha} for DUNE 
and ~\cite{Abe:2016ero} for T2HKK) as well as a precise measurement of 
$\dcp$~\cite{Raut:2017dbh}. Atmospheric neutrino experiments that are 
insensitive to $\dcp$ on their own can exclude the wrong hierarchy 
solution and increase the CP precision in conjunction with the current 
generation of long-baseline experiments~\cite{Ghosh:2015ena}. Finally, 
experiments with small matter effects are ideal for measuring $\dcp$ 
without matter-induced fake CP-violation effects. These include 
muon decay-at-rest experiments~\cite{Agarwalla:2010nn}. Another example is 
\essnusb~\cite{Agarwalla:2014tpa} which aims to measure $\dcp$ at the second oscillation 
maximum, where the CP-sensitivity $\textrm{d}\pmue/\textrm{d}\dcp$ is 
three times more than at the first oscillation maximum. 
Prototype-neutrino-factory experiments like MOMENT can also measure $\dcp$ 
in a synergistic combination with current experiments~\cite{Blennow:2015cmn}. 

As discussed before, we need both neutrino and antineutrino data 
to determine the octant of $\theta_{23}$. DUNE and T2HK which will 
run with both polarities will measure the octant if $\theta_{23}$ 
is not very close to $45^\circ$. Atmospheric neutrino experiments 
like HK, ICAL and PINGU have also been shown to have good sensitivity 
to the octant. It is important to note that the effective atmospheric 
mixing angle measured by experiments is given by
$\sin^2 2\theta_{\mu\mu} = 4 |U_{\mu3}| (1-|U_{\mu3}|^2)$, 
which is a function of both $\theta_{23}$ and $\theta_{13}$. Thus, 
the synergy between long-baseline/atmospheric neutrino experiments 
and reactor neutrino experiments is important in this case, i.e.
a precise measurement of $\theta_{13}$ is crucial. 

Finally we make a few comments on physics beyond the standard 
three-flavour scenario. Addition of extra sterile neutrinos 
introduces more mass-squared difference, mixing angles and CP 
phases that can affect the oscillations and create new degeneracies. 
The existence of charged-current and neutral-current NSIs can also 
dramatically increase the dimension of the oscillation parameter 
space. In such cases, it is important to ask whether 
(a) the presence of these new parameters can affect the measurement 
of the standard oscillation parameters, and (b) our proposed 
experimental facilities can measure these new parameters. 
(While the presence of these large numbers of new parameters can 
seem overwhelming, it is important to remember that the search 
for these parameters is an opportunity to probe physics beyond the 
Standard Model, similar to collider searches.)

Given the large number of new parameters, it is difficult to make 
generalized statements about the degeneracies and their resolution 
in these new physics scenarios. A few specific cases have been studied 
in the literature. For instance, Ref.~\cite{Agarwalla:2016xlg} discusses 
the octant-$\dcp$-$\delta_{14}$ degeneracy in the 3+1 neutrino 
framework, which is hard to resolve even with a combination of neutrino 
and antineutrino data~\cite{Ghosh:2017atj}. Depending on the true values 
of the parameters, experiments like DUNE and T2HKK may be able 
to resolve some of these degeneracies~\cite{Choubey:2017cba}. 

In the case of charged-current NSIs which are already well-bounded, 
experiments with small matter effects like \essnusb\ can probe 
the new parameters~\cite{Blennow:2015nxa}. Neutral-current NSIs on 
the other hand are not constrained strongly at present, and can 
influence our measurements. Combinations of experiments such as 
DUNE with T2HK~\cite{Coloma:2015kiu} and T2HKK~\cite{Liao:2016orc} 
can help in some cases. To conclude, case-by-case studies of the 
various new physics scenarios are needed in order to formulate 
a strategy to deal with the new parameters.

\section{Summary}

We summarize with the help of this flowchart that shows a 
(very simplistic) roadmap for future oscillation studies. 
If the future oscillation data are found 
to be consistent with the standard oscillation framework, one can address 
the degeneracies (if they exist) by using data from different experiments. 
The synergies between various experiments on account of their different 
baseline and energy dependence will help to measure the parameters unambiguously. 
However if new physics effects are found in these data, more detailed studies 
will be required to devise methods to address the problems they will pose.
%--------------------------------------------------------------------------
\begin{figure*}
%\vspace{-5cm}
\begin{center}
\includegraphics[trim=0 0 0 50,clip,width=0.85\textwidth]{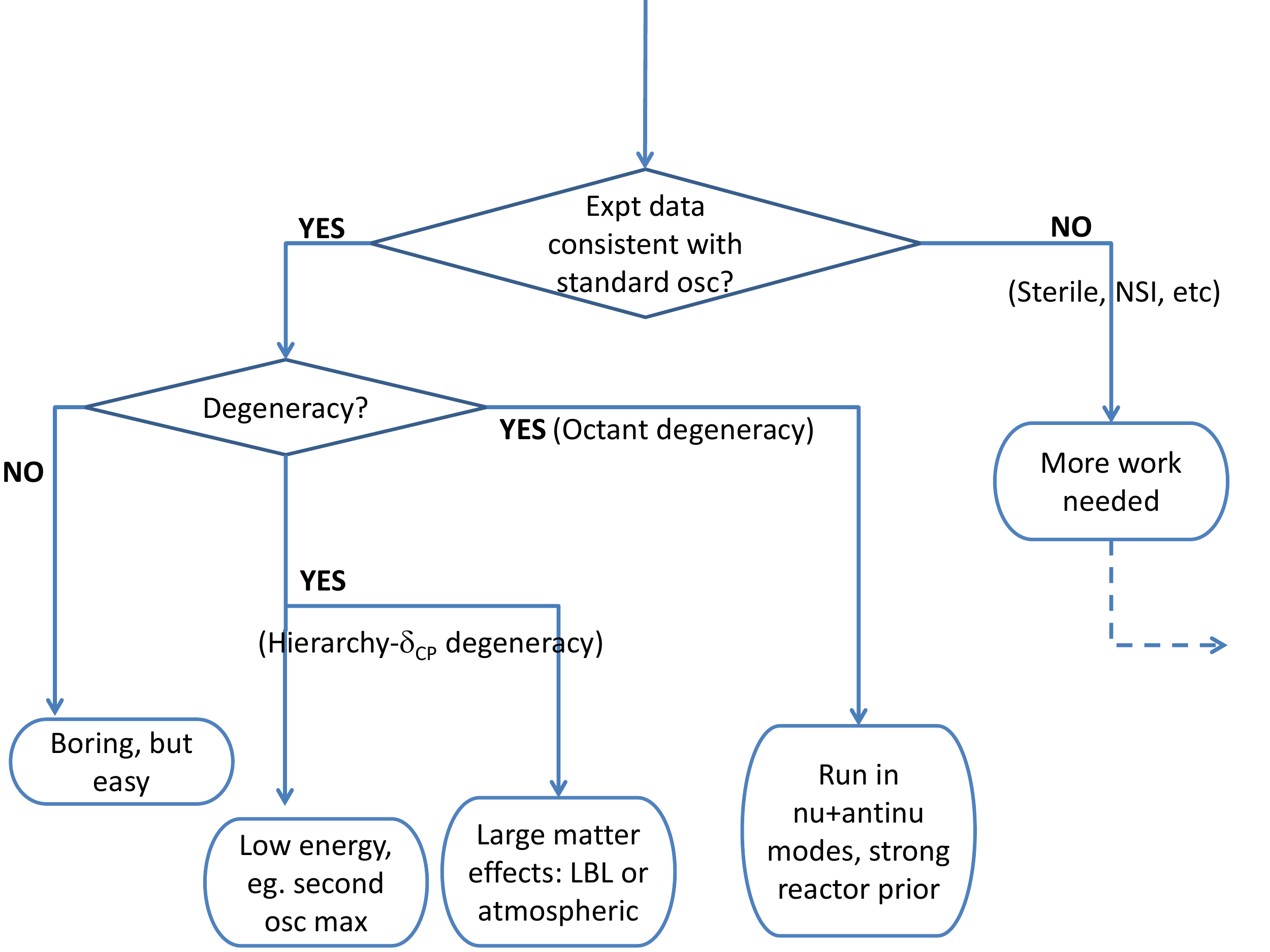}
%\caption{}
\label{fig:flowchart}
\end{center}
\end{figure*}
%--------------------------------------------------------------------------

\bibliographystyle{JHEP}
\bibliography{long-baseline-references}

\end{document}